\documentclass{ifacconf}

\usepackage{amssymb,array,arydshln,color}
\usepackage{cuted}
\usepackage{soul}
\usepackage{flushend}
\usepackage{mathtools}
\usepackage{subcaption}
\usepackage{natbib}
\usepackage[colorinlistoftodos, textwidth=2cm, shadow]{todonotes}

\def\Hc{{\mathcal H}}

\def\Lc{{\mathcal L}}

\def\Rbb{{\mathbb R}}

\def\Sbb{{\mathbb S}}

\def\0{{\bf 0}}

\newcommand{\bitem}{\begin{itemize}}
\newcommand{\eitem}{\end{itemize}}
\newcommand{\btabular}{\begin{tabular}}
\newcommand{\etabular}{\end{tabular}}
\newcommand{\bcenter}{\begin{center}}
\newcommand{\ecenter}{\end{center}}
\newcommand{\bea}{\begin{eqnarray}}
\newcommand{\eea}{\end{eqnarray}}
\newcommand{\bean}{\begin{eqnarray*}}
\newcommand{\eean}{\end{eqnarray*}}
\newcommand{\ba}{\left[ \begin{array}}
\newcommand{\ea}{\\ \end{array} \right]}
\newcommand{\bear}{\begin{array}}
\newcommand{\eear}{\\ \end{array}}

\newcommand{\non}{\nonumber}

\newcommand*{\QEDB}{\hfill\ensuremath{\blacksquare}}%
\newcommand*{\QET}{\hfill\ensuremath{\triangleleft}}
\newcommand{\norm}[1]{\left\lVert#1\right\rVert}

\newcounter{subequation}
\def\beasub{\addtocounter{equation}{+1}
\setcounter{subequation}{\value{equation}}
\setcounter{equation}{0}
\renewcommand{\theequation}{\arabic{subequation}\alph{equation}}
\begin{eqnarray}}
\def\eeasub{\end{eqnarray}
\setcounter{equation}{\value{subequation}}
\renewcommand{\theequation}{\arabic{equation}}}

\def\inf{\operatornamewithlimits{inf\vphantom{p}}}

\newtheorem{Lemma}{Lemma}

\newtheorem{Assumption}{Assumption}
\newtheorem{Remark}{Remark}
\newtheorem{Problem}{Problem}
\newtheorem{Proposition}{Proposition}

\begin{document}
\begin{frontmatter}

\title{Security Metrics for Uncertain Interconnected Systems under Stealthy Data Injection Attacks \thanksref{footnoteinfo}} 

\thanks[footnoteinfo]{This work is supported by the Swedish Research Council under the grant 2021-06316, 2024-00185, the Swedish Foundation for Strategic Research, and by the Knut and Alice Wallenberg Foundation.}

\author[uu]{Anh Tung Nguyen} 
\author[kth]{Sribalaji C. Anand} 
\author[uu]{Andr{\'e} M. H. Teixeira}

\address[uu]{Department of Information Technology, Uppsala University, Uppsala, Sweden (e-mail: \{anh.tung.nguyen, andre.teixeira\}@it.uu.se).}
\address[kth]{KTH Royal Institute of Technology, 
   Stockholm, Sweden (e-mail: scra@kth.se)}

\begin{abstract}                
%
%
%
%
%
%
%
%
This paper quantifies the security of uncertain interconnected systems under stealthy data injection attacks. In particular, we consider a large-scale system composed of a certain subsystem interconnected with an uncertain subsystem, where only the input-output channels are accessible. An adversary is assumed to inject false data to maximize the performance loss of the certain subsystem while remaining undetected. By abstracting the uncertain subsystem as a class of admissible systems satisfying an $\Lc_2$ gain constraint, the worst-case performance loss is obtained as the solution to a convex semi-definite program depending only on the certain subsystem dynamics and such an $\Lc_2$ gain constraint. This solution is proved to serve as an upper bound for the actual worst-case performance loss when the model of the entire system is fully certain. The results are demonstrated through numerical simulations of the power transmission grid spanning Sweden and Northern Denmark.
\end{abstract}


\end{frontmatter}

\section{Introduction}
Networked control systems (NCSs) are a ubiquitous part of many critical infrastructures such as power grids \citep{singh2014stability}, water distribution systems \citep{amin2012cyber}, and transportation systems \citep{park2014high}. Owing to the use of insecure communication protocols in such systems, an increasing number of cyber-attacks are deployed through the communication channels \citep{hemsley2018history}. 
This motivates research on the security of NCSs, particularly from a control theoretic viewpoint over the last decade \citep{dibaji2019systems}. Alongside this progress, one open question is to address the security of uncertain NCSs, which is the objective of this paper.

Some related studies have explored this area. For instance, \cite{park2019stealthy,zhang2022design} construct attack policies when the plant model is uncertain for the adversary. \cite{russo2021poisoning} develops a data-driven attack policy against data-driven controllers.  \cite{harshbarger2020little} studies the role of an uncertain plant model on the stealthiness of an attack. \cite{anand2024risk,anand2023risk} provide a metric to quantify the security against stealthy attacks on uncertain systems. This paper extends the work \citep{anand2024risk} to consider interconnected deterministic and uncertain subsystems.

\begin{figure}[!t]
    \centering
    \includegraphics[width=0.7\linewidth]{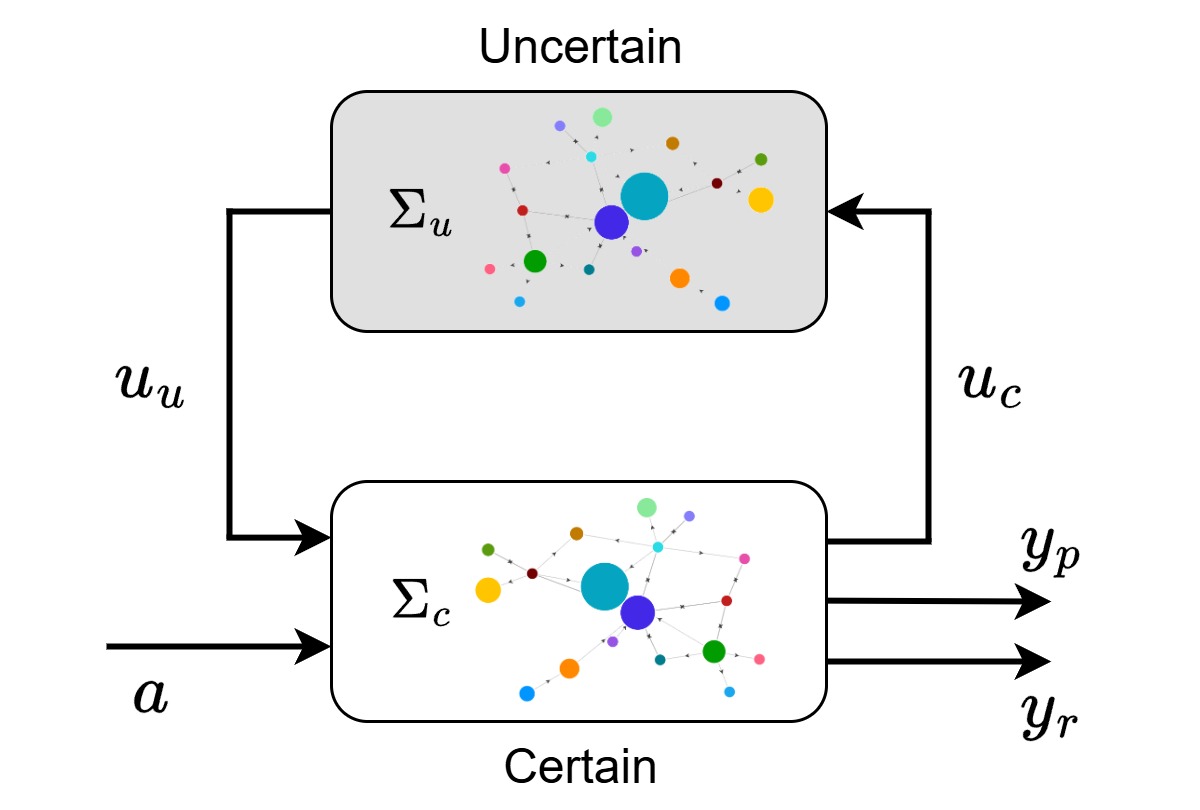}
    \caption{Certain subsystem $\Sigma_c$ interconnected with uncertain subsystem $\Sigma_u$ under stealthy injection attacks $a$.}
    \label{fig:problem}
\end{figure}

In this paper, we consider an interconnected continuous-time NCS, which consists of two subsystems $\Sigma_c$ and $\Sigma_u$ as represented in Fig.~\ref{fig:problem}. The model of $\Sigma_c$ is certain to the adversary and the defender while the model of $\Sigma_u$ is not available to the defender. An adversary is assumed to inject false data into $\Sigma_c$ to increase its performance loss while staying stealthy. Under the above setup, we present the following contributions. 
\begin{enumerate}
    \item An optimization problem is proposed to quantify the worst-case performance loss caused by a stealthy adversary. We observe the problem to be non-convex.
    \item This non-convex optimization problem is shown to be equivalent to a semi-definite program (SDP) problem. 
    \item  We show that the solution to the aforementioned SDP problem serves as an upper bound of the actual performance loss caused by stealthy adversaries. 
\end{enumerate}
The results obtained are demonstrated through numerical simulations of the power transmission grid spanning Sweden and Northern Denmark.
We conclude this section by providing the notation used throughout the paper.

\textbf{Notation:} $\Rbb^n \, (\Rbb^n_{>0})$ and $\Rbb^{n \times m}$ stand for sets of real (positive) $n$-dimensional vectors and real $n$-by-$m$ matrices, respectively; 
the set of $n$-by-$n$ symmetric matrices is denoted as $\Sbb^n$.
We denote $A \preceq 0$ if $A$ is a negative semi-definite matrix.
The space of square-integrable  functions is defined as $\Lc_{2} \triangleq \{f: \Rbb_{>0} \rightarrow \Rbb^n | \norm{f}^2_{\Lc_2 [0,\infty]} < \infty \}$ and the extended space is defined as $\Lc_{2e} \triangleq \{ f: \Rbb_{>0} \rightarrow \Rbb^n | \norm{f}^2_{\Lc_2 [0,H]} < \infty, \forall \, 0 < H < \infty \} $ where $\norm{f}_{\Lc_2 [0,H]}^2 \triangleq \int_{0}^{H} \norm{f(t)}_2^2 \, \text{d}t$.
The notation $\norm{f}^2_{\Lc_2}$  is used  as shorthand for the norm $\norm{f}_{\Lc_2 [0,H]}^2$ if the time horizon $[0,H]$ is clear from the context.

\section{Problem Formulation}\label{sec:problem:formulation}
This section first describes the model of an interconnected system and the attack scenario. The purpose of the adversary and the security problem are formulated in the remainder of the section.

\subsection{System modeling and assumptions}
We consider the security problem in an interconnected system depicted in Fig.~\ref{fig:problem} where the model of the subsystem $\Sigma_c$ is certain and the model of the subsystem $\Sigma_u$ is uncertain. They are described in the following.
\begin{align} \label{model_S1}
    (\Sigma_c): \begin{cases}
        \dot x_c(t) &= A_c x_c(t) + B_c u_u(t) + F_x a(t), \\
        y_p(t) &= C_p x_c(t) + D_p u_u(t) + F_p a(t), \\
        y_r(t) &= C_r x_c(t) + D_r u_u(t) + F_r a(t), \\
        u_c(t) &= C_c x_c(t) + D_c u_u(t) + F_c a(t),
    \end{cases}
\end{align}
where $x_c(t) \in \Rbb^{n_c}$ is the state of $\Sigma_c$, $u_u(t)$ is the input from $\Sigma_u$ to $\Sigma_c$, $a(t)$ is the attack signal, $y_p(t)$ is the performance output, $y_r(t)$ is the residual signal, and $u_c(t)$ is the input from $\Sigma_c$ to $\Sigma_u$. Note that all system matrices are known and are of appropriate dimensions.
Next, the uncertain system $\Sigma_u$ has the following model: 
\begin{align} \label{model_S2}
    (\Sigma_u): \begin{cases}
        \dot x_u(t) &= A_u x_u(t) + B_u u_c(t), \\
        u_u(t) &= C_u x_u(t) + D_u u_c(t),
    \end{cases}
\end{align}
where $x_u(t) \in \Rbb^{n_u}$ is the state of $\Sigma_u$. Note that these system matrices of $\Sigma_u$ are uncertain to the defender but possibly certain to the adversary. Let us employ the following assumptions.
\begin{Assumption}
    [System knowledge] \label{as:knowledge}
    The defender knows the system matrices of $\Sigma_c$ but does not know the system matrices of $\Sigma_u$. 
    \QET
\end{Assumption}
\begin{Assumption}
    [Available data] \label{as:data}
    The defender has access to the online data of $u_c(t)$ and $u_u(t)$. \QET
\end{Assumption}
\begin{Assumption}
    [Stability] \label{as:stability}
    The systems $\Sigma_c$, $\Sigma_u$, and the closed-loop of $\Sigma_c$ and $\Sigma_u$ are internally stable without attacks ($a(t) = 0$).
    \QET
\end{Assumption}
\begin{Assumption}
    [Bounded attack energy] \label{as:attack_bound}
    The energy of the attack signal $a(t)$ in \eqref{model_S1} is bounded. 
    \begin{align}
        \norm{a}_{\Lc_2}^2 \leq E,
    \end{align}
    where $E$ is given. \QET
\end{Assumption}
In practice, the adversary can invest unpredictable time and effort in understanding the system's modeling, which affords them to design optimal attack policies. On the other hand, the defender is not always given the full system knowledge. As a result, Assumption~\ref{as:knowledge} enables us to study more practical systems where some parts of the entire system are not well modeled or hard to model such as nonlinearity or uncertainty. Further, Assumption~\ref{as:data} is employed to support the defender in having some information about the uncertain system $\Sigma_u$, which will be briefly discussed in this paper and plant a seed for future work.   
Assumption~\ref{as:stability} drives our attention to study risk assessment for internally stable systems since states of unstable systems diverge before an adversary appears. Thus, unstable systems should prioritize stabilizing controllers rather than risk assessment. We assume that unstable systems are equipped with a stabilizing controller, resulting in an internally stable closed-loop system. More details on how system matrices of $\Sigma_c$ and $\Sigma_u$ incorporate stabilizing controllers and anomaly detectors can be found in \cite{teixeira2021security}.
As a consequence, Assumption~\ref{as:stability} allows us to assume that the states of $\Sigma_c$ and $\Sigma_u$ are at its equilibrium $x_e = 0$ before being affected by the adversary, i.e.,
\begin{align}
    x_c(0) = 0,~~x_u(0) = 0. 
\end{align}
Assumption~\ref{as:attack_bound} enables us to consider that the attack signal vanishes when time goes to infinity, resulting in 
\begin{align}
    x_c(\infty) = 0,~~x_u(\infty) = 0. 
\end{align}
\vspace{-0.5cm}
\subsection{Adversarial purpose \& security metric}
We define that the adversary is detected if the energy of the residual signal $y_r$ over a time horizon $[0,H]$ crosses a given alarm threshold $\delta > 0$:
\begin{align}
    \norm{y_r}^2_{\Lc_2[0, H]} > \delta.
\end{align}
Adversaries, who are detected, can be mitigated quickly and efficiently. Therefore, we are interested in studying stealthy attacks that always guarantee the energy of the residual signal $y_r$ under the alarm threshold $\delta$, i.e.,
\begin{align}
    \norm{y_r}^2_{\Lc_2[0, H]} \leq \delta.
\end{align}
At the same time, the adversary aims to design an attack signal that maximizes the performance output of the system $\Sigma_c$. The two aforementioned adversarial purposes (stealthiness and maximum performance loss) are formulated in the following optimization problem:
\begin{align}
    Q \triangleq \sup_{a} ~~& \norm{y_p}_{\Lc_2}^2 \label{Q_sup} \\
    \text{s.t.}~~& 
    \eqref{model_S1}, \eqref{model_S2}, x_c(0) = 0, x_u(0) = 0, \non \\
    & x_c(\infty) = 0, x_u(\infty) = 0, \non \\
    &
    \norm{y_r}^2_{\Lc_2} \leq \delta, 
    \norm{a}^2_{\Lc_2} \leq E. 
    \non
\end{align}
The impact \eqref{Q_sup} serves as a security metric since it quantifies the maximum risk of a given system subject to stealthy injection attacks \citep{teixeira2015strategic}.
This metric was utilized for designing defense and mitigation strategies in our previous work such as \cite{nguyen2024security} for certain system models and \cite{anand2024risk} for system models with probabilistic parameter uncertainty. 
Alongside this progress, this paper extends the risk management under uncertainty by investigating the following problem.
\begin{Problem} 
    \label{pb:security_metric}
    Given the interconnected system \eqref{model_S1}-\eqref{model_S2} depicted in Fig.~\ref{fig:problem} and Assumptions~\ref{as:knowledge}-\ref{as:attack_bound}, quantify the worst-case impact of stealthy attacks \eqref{Q_sup}.
\end{Problem}

\section{Quantifying security metric}\label{sec:main}
In this section, we first provide computation for \eqref{Q_sup} when the models of $\Sigma_c$ and $\Sigma_u$ are fully certain. This value serves as the ground-truth attack impact for the comparison with a solution to Problem~\ref{pb:security_metric}, which is dealt with in the remainder of the section.
\subsection{Full system knowledge}
In this section, we provide the quantification of \eqref{Q_sup} with full system knowledge, i.e., we assume that the system matrices of $\Sigma_c$ and $\Sigma_u$ are certain for now.
\begin{Lemma}
    [Output-to-output gain] \label{lem:Q_quantification} 
    Given the system \eqref{model_S1}-\eqref{model_S2} under stealthy data injection attacks, the worst-case impact of stealthy data injection attacks \eqref{Q_sup} is equivalent to the following optimization problem:
    \begin{align}
        Q = \min_{P, \gamma, \psi}& ~~ \gamma \delta + \psi E \label{Q_dual_min} \\
    \text{s.t.}&~~
    P \in \Sbb^{n_c+n_u}, \gamma > 0, \psi > 0, \non \\ 
     \dot V_o(x_c(t), x_u(t)) &\leq s_o(y_r(t),a(t),y_p(t)), \forall (a(t), x_c(t), x_u(t)), \non
    \end{align}
    where $V_o(x_c(t), x_u(t)) \triangleq [x_c(t)^\top, x_u(t)^\top] P [x_c(t)^\top, x_u(t)^\top]^\top$ is a storage function and $s_o(y_r(t),a(t),y_p(t)) \triangleq \gamma \norm{y_r(t)}_{2}^2 + \psi \norm{a(t)}_{2}^2 - \norm{y_p(t)}_{2}^2$ is a supply rate. \QET
\end{Lemma}
\begin{pf}
Since the attack energy is bounded $a \in \Lc_2$, it follows from \cite{anand2024risk}, that \eqref{Q_sup} is equivalent to its dual form given as:
    \begin{align} 
        \inf_{\gamma, \psi}& \bigg[ \sup_{a} \big[ \norm{y_p}^2_{\Lc_2} + \gamma \big( \delta - \norm{y_r}^2_{\Lc_2} \big) + \psi \big( E - \norm{a}^2_{\Lc_2} \big) \big] \bigg] \non \\
        \text{s.t.}&~ \eqref{model_S1}-\eqref{model_S2}, x_c(0) = 0, x_u(0) = 0, \non \\
        &~x_c(\infty) = 0, x_u(\infty) = 0,
        \gamma > 0, \psi > 0. \label{Q_sup_dual_lm1}
    \end{align}
    The dual form \eqref{Q_sup_dual_lm1} is bounded if, and only if, the following condition holds
    \begin{align}
        \norm{y_p}^2_{\Lc_2} - \gamma \norm{y_r}^2_{\Lc_2} - \psi \norm{a}^2_{\Lc_2} \leq 0, \label{Q_sup_dual_cond}
    \end{align}
    for all cyclic trajectories with $x_c(0) = x_c(\infty) = 0$ and $x_u(0) = x_u(\infty) = 0$. The inequality \eqref{Q_sup_dual_cond} can be rewritten as: $ \int_{0}^\infty s_o(y_r(t),a(t),y_p(t)) \text{d}t \geq 0 = V_o(x_c(\infty), x_u(\infty)) - V_o(x_c(0), x_u(0))$, which is equivalent to the system \eqref{model_S1}-\eqref{model_S2} being cyclo-dissipative \citep{moylan2014dissipative}. Taking time-derivative both sides of this dissipation inequality gives us the last constraint of \eqref{Q_dual_min} with $P$ being symmetric. 
    \QEDB
\end{pf}
The last constraint of \eqref{Q_dual_min} can be equivalently rewritten in the following linear matrix inequality (LMI) form:
\begin{align}
    \ba{cc}
    \bar A^\top P + P \bar A ~&~ P \bar F \\
    \bar F^\top P ~&~ 0 
    \ea ~\preceq~& \gamma \ba{c} \bar C_r^\top \\ \bar F_r^\top \ea 
    \ba{c} \bar C_r^\top \\ \bar F_r^\top \ea^\top + \psi \ba{c}
    0 \\ I \ea \ba{c}
    0 \\ I \ea^\top \non \\
    &~~~~ - \ba{c} \bar C_p^\top \\ \bar F_p^\top \ea 
    \ba{c} \bar C_p^\top \\ \bar F_p^\top \ea^\top,
    \label{lmi_constraint}
\end{align}
where all the system matrices can be found in Appendix~\eqref{matrix_agg}. Hence, \eqref{Q_dual_min} becomes a tractable SDP problem with constraints in the form of LMIs \eqref{lmi_constraint}.

It is worth noting that the problem \eqref{Q_dual_min} with the LMI constraint \eqref{lmi_constraint} can only be solved if all the system matrices are certain, which does not hold true under Assumption~\ref{as:knowledge}. In the next subsection, we provide the quantification of \eqref{Q_dual_min} when the system matrices of $\Sigma_u$ are uncertain. 
\subsection{Limited system knowledge}
Although the system matrices of $\Sigma_u$ are uncertain under Assumption~\ref{as:knowledge}, the input-output data of $\Sigma_u$ is accessible by employing Assumption~\ref{as:data}. This assumption enables us to use the following input-output relationship:
\begin{align}
    \norm{u_u}_{\Lc_2}^2 \leq \gamma_u \norm{u_c}_{\Lc_2}^2,
    \label{G_2_bound}
\end{align}
where $\gamma_u$ is given. Note that $\gamma_u$ is the $\Hc_\infty$ norm of the system \eqref{model_S2}. How to obtain $\gamma_u$ will be discussed in Section~\ref{sec:compute_g2}. By abstracting the uncertain $\Sigma_u$ as a class of admissible systems satisfying \eqref{G_2_bound}, we propose the following proxy worst-case impact of stealthy data injection attacks:
\begin{align}
    \hat Q \triangleq \sup_{a, \, u_u \in \Lc_2} ~~& \norm{y_p}_{\Lc_2}^2 \label{Q_hat_sup} \\
    \text{s.t.}~~& 
    \eqref{model_S1}, 
    x_c(0) = 0, x_c(\infty) = 0, 
    \non \\
    &
    \norm{y_r}^2_{\Lc_2} \leq \delta, ~
    \norm{a}^2_{\Lc_2} \leq E, \non \\
    &
    \norm{u_u}_{\Lc_2}^2 \leq \gamma_u \norm{u_c}_{\Lc_2}^2, 
    \non
\end{align}
where the uncertain system modeling \eqref{model_S2} in \eqref{Q_sup} is replaced with the input-output relationship \eqref{G_2_bound}. The following theorem introduces a method to compute \eqref{Q_sup} using the cyclo-dissipative system theory \citep{moylan2014dissipative}.
\begin{thm}
    [Proxy output-to-output gain] \label{th:security_metric}
    Given the system \eqref{model_S1} under stealthy data injection attacks and interconnected with the uncertain system \eqref{model_S2}, the proxy output-to-output gain security metric \eqref{Q_hat_sup} is equivalent to the following optimization problem:
    \begin{align}
        \hat Q = \min_{P_c, \gamma, \psi, \theta} &~~ \gamma \delta + \psi E \label{Q_hat_min} \\
    \text{s.t.}~~&\gamma > 0, \psi > 0, \theta > 0, P_c \in \Sbb^{n_c}, \non \\
    &\dot V_c(x_c(t)) \leq s_c(y_r(t),a(t),u_u(t),y_p(t),u_c(t)), \non \\
    &\hspace{2.5cm}
    \forall (a(t),x_c(t),u_u(t), u_c(t)), \non 
    \end{align}
    where $V_c(x_c(t)) \triangleq x_c(t)^\top P_c x_c(t)$ is a storage function and $s_c(y_r(t),a(t),u_u(t),y_p(t),u_c(t)) \triangleq \gamma \norm{y_r(t)}^2_{2} + \psi \norm{a(t)}^2_{2} + \theta \norm{u_u(t)}_{2}^2 - \norm{y_p(t)}_{2}^2 - \theta \gamma_u  \norm{u_c(t)}_{2}^2$ is a supply rate.
    \QET
\end{thm}
\begin{pf}
Since $a, u_u \in \Lc_2$, \eqref{Q_hat_sup} is equivalent to its dual form \citep{anand2024risk} given as:
    \begin{align}
    \inf_{\gamma, \psi, \theta} \bigg[ ~ \sup_{a, u_u} &\big[  \norm{y_p}_{\Lc_2}^2 + \gamma (\delta - \norm{y_r}^2_{\Lc_2}) + \psi ( E -  \norm{a}^2_{\Lc_2} ) \non \\
    &~~~~~~
    + \theta (\gamma_u \norm{u_c}_{\Lc_2}^2 - \norm{u_u}_{\Lc_2}^2) \big]~ \bigg] \label{Q_sup_dual}
    \\ 
    \text{s.t.}~\eqref{model_S1},& 
     ~ x_c(0) = 0, x_c(\infty) = 0, 
    \gamma > 0, \psi > 0, \theta > 0. \non
    \end{align}
    The dual form is bounded if, and only if, 
    \begin{align}
        &\norm{y_p}_{\Lc_2}^2 - \gamma \norm{y_r}^2_{\Lc_2} - \psi \norm{a}^2_{\Lc_2} + \theta \gamma_u  \norm{u_c}_{\Lc_2}^2   \non \\ 
        &\hspace{5.0cm}
        - \theta \norm{u_u}_{\Lc_2}^2 \leq 0, 
         \label{Q_hat_dual_cond}
    \end{align}
    for all cyclic trajectories with $x_c(0) = x_c(\infty) = 0$,
    which is equivalent the system \eqref{model_S1} being cyclo-dissipative with the storage function $V_c(x_c(t))$
    and the supply rate $s_c(y_r(t),a(t),u_u(t),y_p(t),u_c(t))$.
    By employing the cyclo-dissipative system theory \citep{moylan2014dissipative}, \eqref{Q_hat_dual_cond} is equivalent to  
    the last constraint of \eqref{Q_hat_min} with symmetric $P_c$. 
    \QEDB
\end{pf}

The last constraint of \eqref{Q_hat_min} can be rewritten as follows: \( [x_c(t)^\top, u_u(t)^\top, a(t)^\top] \times R_c(\Sigma_c, P_c, \gamma_u, \gamma, \psi, \theta) \times \) \\
\([ x_c(t)^\top, u_u(t)^\top, a(t)^\top ]^\top \leq 0 \),
which holds for any trajectory $(a(t),x_c(t),u_u(t), u_c(t))$ if the following LMI constraint holds true:
\begin{align}
    R_c(\Sigma_c, P_c, \gamma_u, \gamma, \psi, \theta)  \preceq 0, \label{Q_hat_lmi_constraint}
\end{align}
where $R_c(\Sigma_c, P_c, \gamma_u, \gamma, \psi, \theta)$ is given in Appendix~\eqref{matrix_R1}. By replacing the last constraint of \eqref{Q_hat_min} with the LMI constraint \eqref{Q_hat_lmi_constraint}, the proxy worst-case impact of stealthy data injection attacks \eqref{Q_hat_sup} can be equivalently computed by solving the resulting SDP problem. It is worth noting that all the system parameters in \eqref{Q_hat_lmi_constraint} are certain, thus solving \eqref{Q_hat_min} with \eqref{Q_hat_lmi_constraint} is possible. 
The following theorem establishes the connection between the worst-case impact of stealthy data injection attacks \eqref{Q_sup} and its proxy \eqref{Q_hat_sup}.
\begin{thm}
    [Upper bound] \label{th:upper_bound} Consider $Q$ in \eqref{Q_sup} and $\hat Q$ in \eqref{Q_hat_sup} and their computations provided in Lemma~\ref{lem:Q_quantification} and Theorem~\ref{th:security_metric}, respectively, one obtains
    \begin{align}
        Q \leq \hat Q. \label{Q_upperbound}
    \end{align}
\end{thm}
\begin{pf}
Let us recall the storage function $V_o(x_c(t), x_u(t))$, the supply rate $s_o(y_r(t),a(t),y_p(t))$ in Lemma~\ref{lem:Q_quantification}, and the last constraint of \eqref{Q_dual_min} which is
\begin{align}
    \dot V_o(x_c(t), x_u(t)) - &s_o(y_r(t),a(t),y_p(t)) \leq 0,  \non \\
    &\hspace{1.5cm}
    \forall (a(t), x_c(t), x_u(t)). \label{Vtst_dissipative}
\end{align}
Let us also recall the storage function $V_c(x_c(t))$, the supply rate $s_c(y_r(t),a(t),u_u(t),y_p(t),u_c(t))$ in Theorem~\ref{th:security_metric}, and the last constraint of \eqref{Q_hat_min} which is
\begin{align}
    \dot V_c(x_c(t)) - &s_c(y_r(t),a(t),u_u(t),y_p(t),u_c(t)) \leq 0, \non \\
    &\hspace{1.5cm}
    \forall (a(t), x_c(t), u_u(t), u_c(t)). \label{V1s1_dissipative}
\end{align}
Based on the cyclo-dissipative system theory \citep{moylan2014dissipative}, the input-output relationship \eqref{G_2_bound} can be equivalently rewritten as follows:
\begin{align}
    \dot V_u(x_u(t)) - s_u(u_c(t),u_u(t)) \leq 0, ~\forall (u_c(t), x_u(t)), \label{V2s2_dissipative}
\end{align}
where $V_u(x_u(t)) \triangleq x_u(t)^\top P_u x_u(t)$ is the storage function and $s_u(u_c(t),u_u(t)) \triangleq \gamma_u \norm{u_c(t)}^2_{2} - \norm{u_u(t)}^2_{2}$ is the supply rate. Next, we consider the following storage function 
\begin{align}
    V_a(x_c(t), x_u(t)) = V_c(x_c(t)) + \theta V_u(x_u(t)), ~~ (\theta > 0). \label{Vc_def}
\end{align}
Let us construct an optimization problem similar to \eqref{Q_hat_min}, by replacing the storage function $V_c(x_c(t))$ and the supply rate $s_c(y_r(t),a(t),u_u(t),y_p(t),u_c(t))$ with the storage function \eqref{Vc_def} and the supply rate $s_o(y_r(t),a(t),y_p(t))$ in Lemma~\ref{lem:Q_quantification}, respectively, while the objective function remains unchanged as follows:
\begin{align}
    \underline Q \triangleq &\min_{P_c, P_u, \gamma, \psi, \theta} \gamma \delta + \psi E \label{Q_underline_min} \\
    \text{s.t.}&~ \gamma > 0, \psi > 0, \theta > 0, P_c \in \Sbb^{n_c}, P_u \in \Sbb^{n_u}, \non \\
    & 
    \dot V_a(x_c(t), x_u(t)) - s_o(y_r(t),a(t),y_p(t)) \leq 0, \non \\
    &\hspace{4cm}
    \forall (a(t), x_c(t), x_u(t)). \non 
\end{align} 
For all cyclic trajectories $(a(t), x_c(t), x_u(t))$, if there exists a tuple $(P_c, P_u, \gamma, \psi, \theta)$ satisfying 
\eqref{V1s1_dissipative} and \eqref{V2s2_dissipative}, then $(P_c, P_u, \gamma, \psi, \theta)$ also satisfies the following inequality
\begin{align}
    &\dot V_a(x_c(t), x_u(t)) = \dot V_c(x_c(t)) + \theta \dot V_u(x_u(t)) \non \\
    &
    \leq s_c(y_r(t),a(t),u_u(t),y_p(t),u_c(t)) + \theta s_u(u_c(t),u_u(t)) \non \\
    &
    = s_o(y_r(t),a(t),y_p(t)), \label{Vcst_dissipative}
\end{align}
On the other hand, the two constraints \eqref{V1s1_dissipative} and \eqref{V2s2_dissipative} are embedded in the last constraint of \eqref{Q_hat_min} while the constraint \eqref{Vcst_dissipative} is employed in \eqref{Q_underline_min}. As a result, the feasible set $(P_c, P_u, \gamma, \psi, \theta)$  of \eqref{V1s1_dissipative} and \eqref{V2s2_dissipative} (and thus of \eqref{Q_hat_min}) is contained by the feasible set of \eqref{Q_underline_min}. Consequently, the solution to \eqref{Q_underline_min} serves as the lower bound of the solution to \eqref{Q_hat_min}, i.e.,
\begin{align}
    \underline Q \leq \hat Q. \label{Q_lower_bound}
\end{align}

From another view on the structure of \eqref{Q_underline_min}, the last constraint of \eqref{Q_underline_min} is constructed by restricting the matrix $P$ of the storage function defined in \eqref{Q_dual_min} as a block diagonal matrix, i.e., $P = \text{blkdiag}(P_c, \theta P_u)$. Further, \eqref{Q_underline_min} and \eqref{Q_dual_min} have the same objective function, naturally resulting in that \eqref{Q_underline_min} is the upper bound of \eqref{Q_dual_min}, i.e., 
\begin{align}
    Q \leq \underline Q. \label{Q_lower_bound2}
\end{align}
Finally,
from \eqref{Q_lower_bound} and \eqref{Q_lower_bound2}, one obtains \eqref{Q_upperbound}. \QEDB
\end{pf}
\begin{Remark}
    The result presented in Theorem~\ref{th:security_metric} enables us to compute a proxy of the worst-case impact of stealthy data injection attacks \eqref{Q_sup} by solving \eqref{Q_hat_min} where the last constraint is replaced with the LMI \eqref{Q_hat_lmi_constraint}. Further, the result of Theorem~\ref{th:upper_bound} guarantees that the solution to the proxy problem \eqref{Q_hat_min} serves as an upper bound of the ground-truth impact \eqref{Q_sup}. This upper bound is beneficial to a defender in further dealing with security allocation problems \citep{anand2023risk,nguyen2024security} under uncertainty, which is left to future work.
\end{Remark}

\section{Discussion on estimating $\gamma_u$}
\label{sec:compute_g2}
In this section, we recall several existing methods in the literature on how to obtain $\gamma_u$ in \eqref{G_2_bound}. Note that we do not provide a contribution, but rather adopt two practical solutions to estimating $\gamma_u$ in \eqref{G_2_bound}. Further theoretical analysis will be left to future work.
\subsection{Model-based estimation}
We first provide an estimation of $\gamma_u$ in \eqref{G_2_bound} based on system matrices \eqref{model_S2} in the following proposition.
\begin{Proposition}
    [Model-based estimation of $\gamma_u$] \label{prop:model_based_gamma2}
    Given the system \eqref{model_S2} and the relationship \eqref{G_2_bound}, the value of $\gamma_u$ in \eqref{G_2_bound} is computed by solving the following SDP problem:
    \begin{align}
    &\min_{P_u \in \Sbb^{n_u}, \,  \gamma_u > 0} ~~  \gamma_u \label{model_based_gamma2} \\
    \text{s.t.}~~~&
    \ba{cc}
    A_u^\top P_u + P_u A_u ~&~ P_u B_u \\ 
    B_u^\top P_u & -\gamma_u I 
    \ea + \ba{c}
    C_u^\top \\ D_u^\top 
    \ea 
    \ba{c}
    C_u^\top \\ D_u^\top 
    \ea^\top  \preceq 0.
    \tag*{\QET} 
    \end{align}
\end{Proposition}
\begin{pf}
    The proof is straightforward by applying the cyclo-dissipative system theory once again with the storage function $V_u(x_u(t)) \triangleq x_u(t)^\top P_u x_u(t)$ and the supply rate $s_u(u_c(t),u_u(t)) \triangleq \gamma_u \norm{u_c}^2_{\Lc_2} - \norm{u_u}^2_{\Lc_2}$. \QEDB
\end{pf}
It is worth noting that $\gamma_u$ is computed by solving \eqref{model_based_gamma2} if all the system matrices \eqref{model_S2} are certain, which does not hold true under Assumption~\ref{as:knowledge}. 
However, the value of $\gamma_u$ obtained by solving \eqref{model_based_gamma2} can be used as the ground truth for the data-based method that considers both Assumptions~\ref{as:knowledge}-\ref{as:data} in the next subsection.
\begin{Remark}
    In the case of full system knowledge, the system can also be decomposed into two parts where $\Sigma_u$ is less important and does not contribute to the performance specification. Then, the method computing $\gamma_u$ in Proposition~\ref{prop:model_based_gamma2} is used to proceed with the security quantification in Theorem~\ref{th:security_metric}. This decomposition definitely reduces the demand for computational resources.
\end{Remark}
\subsection{Extremum-seeking estimation}
\begin{figure}[!t]
    \centering
    \includegraphics[width=0.85\linewidth]{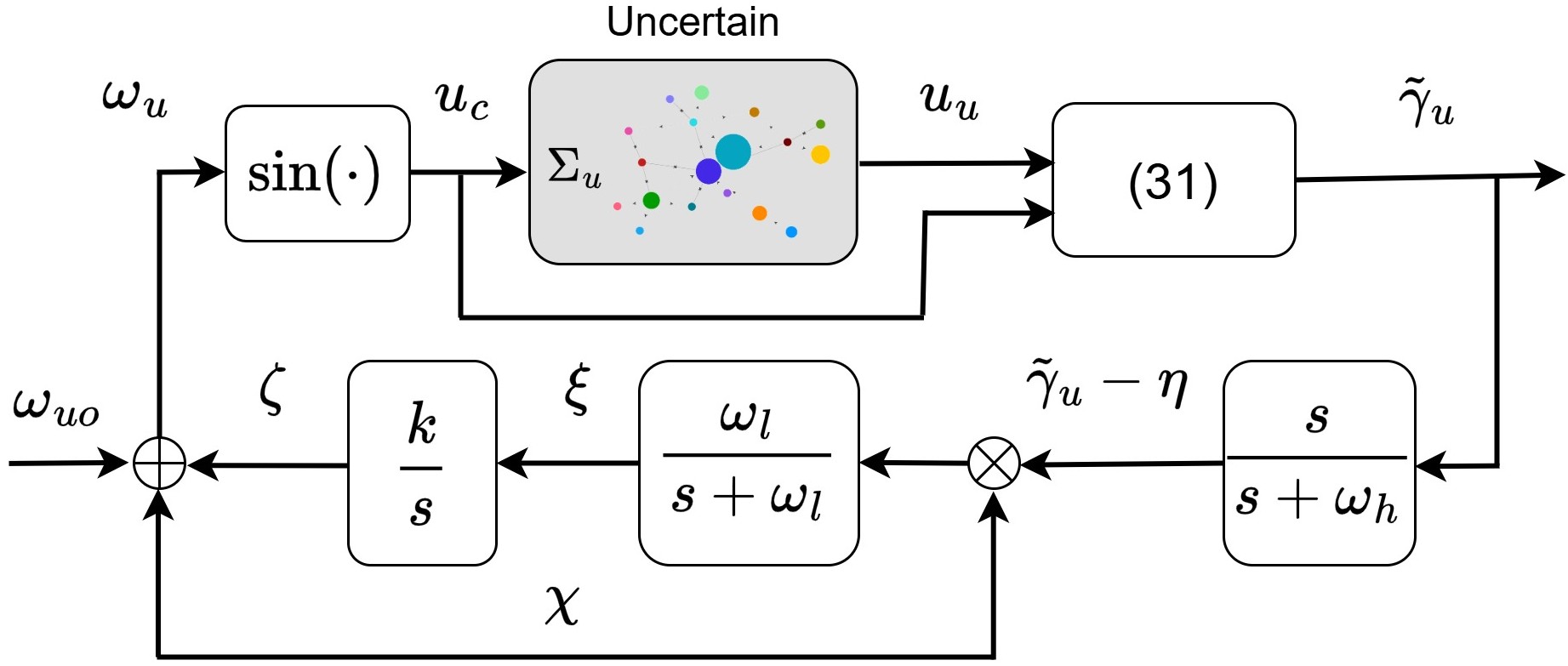}
    \caption{Diagram of extremum-seeking estimation of $\gamma_u$.}
    \label{fig:XseekingSecurity}
\end{figure}

\begin{figure}[!h]
    \centering
    \begin{subfigure}
        {0.48\linewidth}
        \includegraphics[width=1\textwidth]{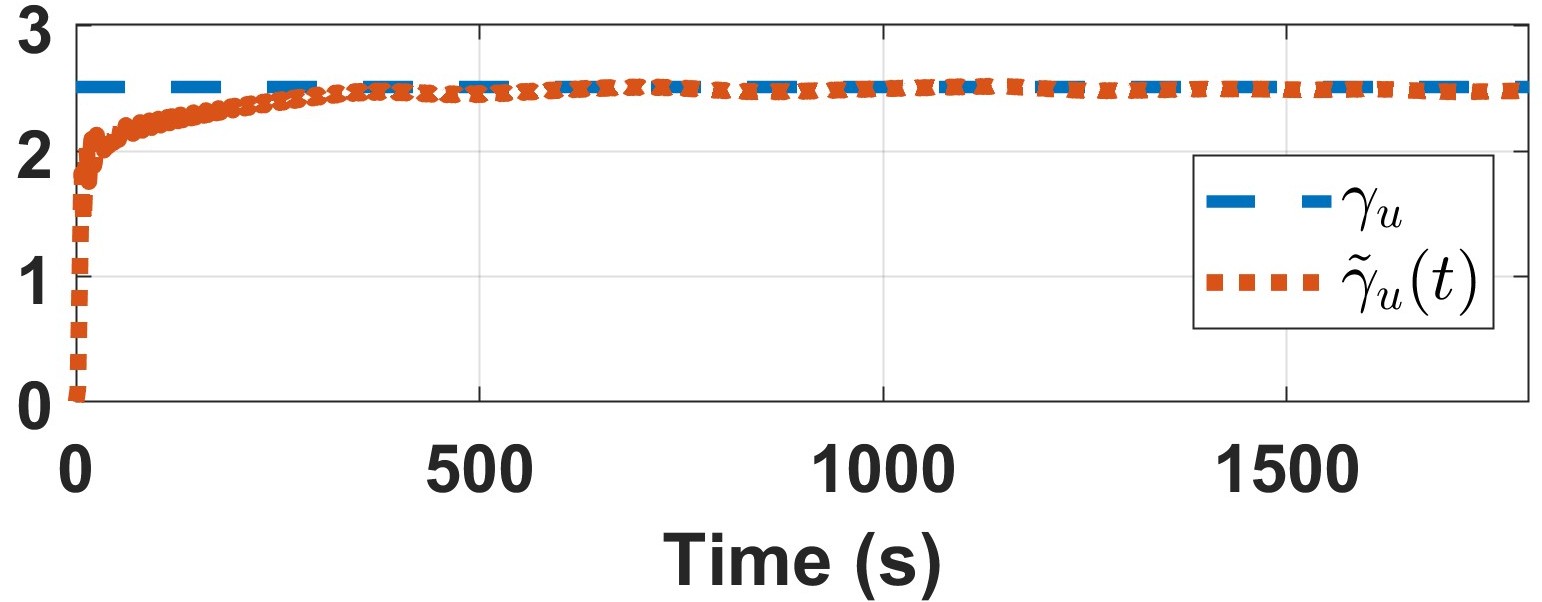}
    \end{subfigure}
    \begin{subfigure}
        {0.48\linewidth}
        \includegraphics[width=1\textwidth]{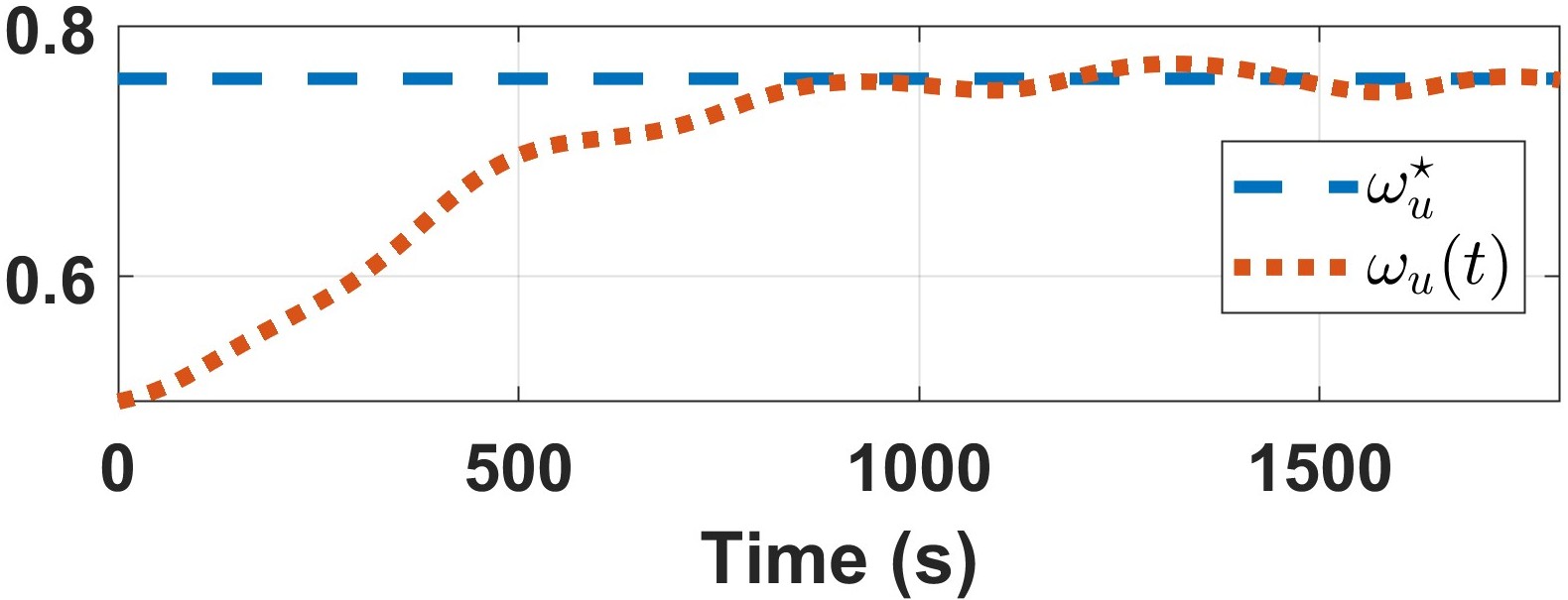}
    \end{subfigure}
    \caption{An example of extremum-seeking estimation of $\gamma_u$.}
    \label{fig:eg_hinf}
\end{figure}

Assumption~\ref{as:data} enables us to inject input $u_c$ into the uncertain system $\Sigma_u$ and measure the output $u_u$ with the aim of approximating the value of $\gamma_u$. From a system theory perspective, the maximum value of $\gamma_u$ corresponds to the maximum singular value of $\Sigma_u$, a fixed input frequency (says $\omega_u^\star$).
By exploiting this property,
we adopt the extremum-seeking feedback \citep{krstic2000stability} to approximate $\gamma_u$ in \eqref{G_2_bound} in the following procedure for a single-input-single-output system: 
\begin{align}
    u_c(t) & = \sin( t \omega_u(t)), \label{esc_y1} \\
    \omega_u(t) & = \omega_{uo} + \chi(t) + \zeta(t), \,
    \chi(t)  = \alpha_p \sin(\omega_p t + \phi_p), \label{esc_chi_i} \\
    \tilde \gamma_u(t) &= {\norm{u_u}^2_{\Lc_2[0, t]}}/{\norm{u_c}^2_{\Lc_2[0, t]}}, \label{esc_g2} \\
    \dot \eta(t) & = \!\! - \omega_h \eta(t) + \omega_h \tilde \gamma_u(t), \label{esc_hpf} \\
     \dot \xi(t) &= \!\! - \omega_l \xi(t) + \omega_l (\tilde \gamma_u(t) - \eta(t)) \chi(t), 
    \dot \zeta(t) = k \xi(t), \label{esc_esd}
\end{align}
where $\omega_{uo}$ is a base frequency, $\chi(t)$ is the harmonic perturbation, $\zeta(t)$ is the extremum seeking direction, $\omega_u(t)$ is the time-varying approximate of $\omega_u^\star$, $\tilde \gamma_u(t)$ is the time-varying approximate of $\gamma_u$ in \eqref{G_2_bound}, $\eta(t)$ is the output of a high pass filter, and $\xi(t)$ is the output of a low pass filter. Parameters $\alpha_p$, $\omega_p$, and $\phi_p$ of the harmonic perturbation, $\omega_h$ of the high pass filter, $\omega_l$ of the low pass filter, and $k$ of the extremum-seeking direction are designed. 
How to choose these parameters to guarantee stability can be found in \cite{krstic2000stability}. 
It is worth noting that the system parameters $(A_u, B_u, C_u, D_u)$ in \eqref{model_S2} are not needed since we only need to measure the output $u_u$ to proceed with \eqref{esc_y1}-\eqref{esc_esd}, which is depicted in Fig.~\ref{fig:XseekingSecurity}. An example of running the procedure \eqref{esc_y1}-\eqref{esc_esd} is reported in Fig.~\ref{fig:eg_hinf} where the extremum-seeking estimate $\tilde \gamma_u(t)$ converges to the model-based $\gamma_u$ computed by solving \eqref{model_based_gamma2}. In the following section, we provide more comprehensive simulations to validate the obtained results.

\vspace{-0.3cm}
\section{Simulation results}\label{sec:NE}
\begin{figure}[!t]
    \centering
    \includegraphics[width=0.7\linewidth]{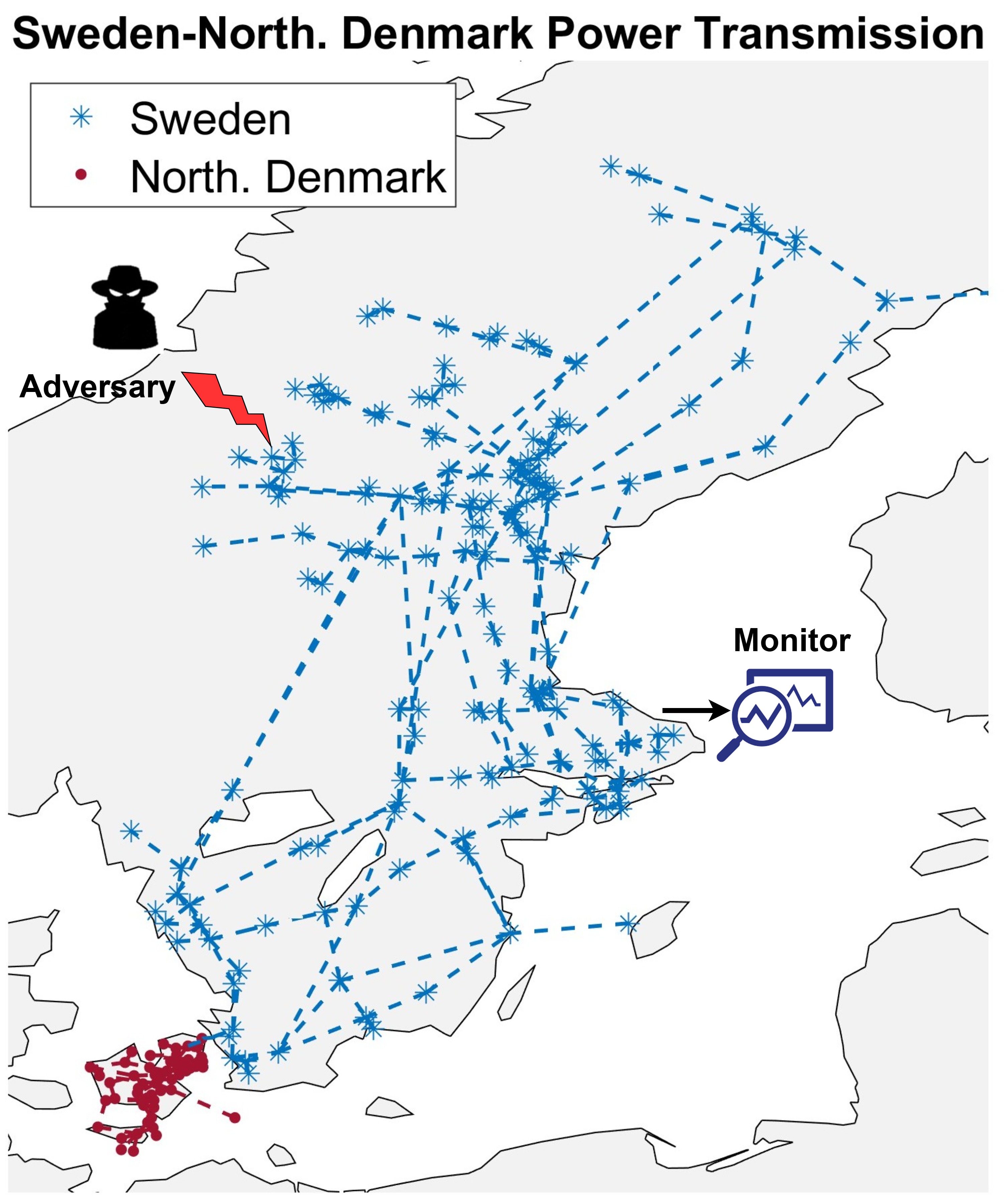}
    \caption{The Sweden-Northern Denmark power transmission grid under stealthy injection attacks where the model of the Northern Denmark grid is uncertain.}
    \label{fig:SEgrid}
\end{figure}
\vspace{-0.3cm}
In this section, we validate the results obtained in the previous sections using the Sweden-Northern Denmark power transmission grid\footnote{Simulation available at: \textit{https://tinyurl.com/UncInt-SecMetric}}, depicted in Fig.~\ref{fig:SEgrid}. In this example, the Sweden grid (195 buses) is considered as the certain subsystem $\Sigma_c$ while the Northern Denmark grid (65 buses) serves as the uncertain subsystem $\Sigma_u$ in Fig.~\ref{fig:problem}. Each bus is modeled by the second-order swing equation \citep{tegling2018fundamental} and the grid parameters are provided in \cite{N490data}.
The authors are thankful to the Division of Signals and Systems, Uppsala University, for providing the computing facility to perform the numerical simulations.
The simulations were conducted in Matlab 2020b on a server that has a configuration: 48 CPU Intel Xenon Gold 6248R 3.0 GHz and 768 Gb RAM DDR4. 

In our simulations, the attack scenario involves an adversary targeting multiple buses while specific buses are monitored. The monitor nodes remain fixed in major cities such as Stockholm and Gothenburg, while the attack nodes are randomly selected. We evaluate the results across three scenarios, where the number of attack buses and monitor buses varies between $2$, $4$, and $6$. For each scenario, 20 attack cases are randomly generated.

\begin{figure} [!h]
    \centering
    \begin{subfigure}
        {0.45\linewidth}
        \includegraphics[width=1\textwidth]{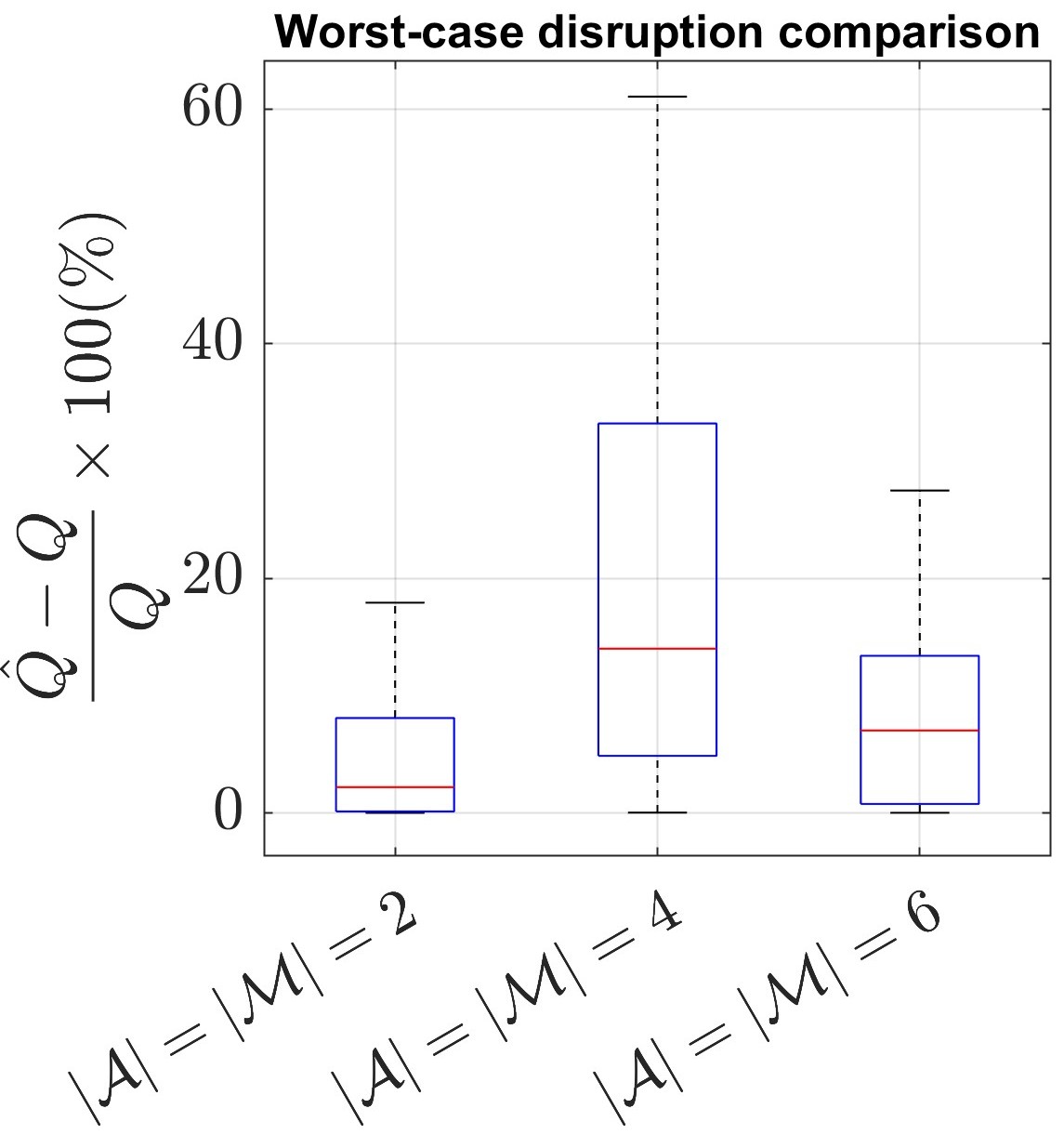}
        \caption{}
        \label{fig:SEgrid_proxy}
    \end{subfigure}
    \begin{subfigure}
        {0.45\linewidth}
        \includegraphics[width=1\textwidth]{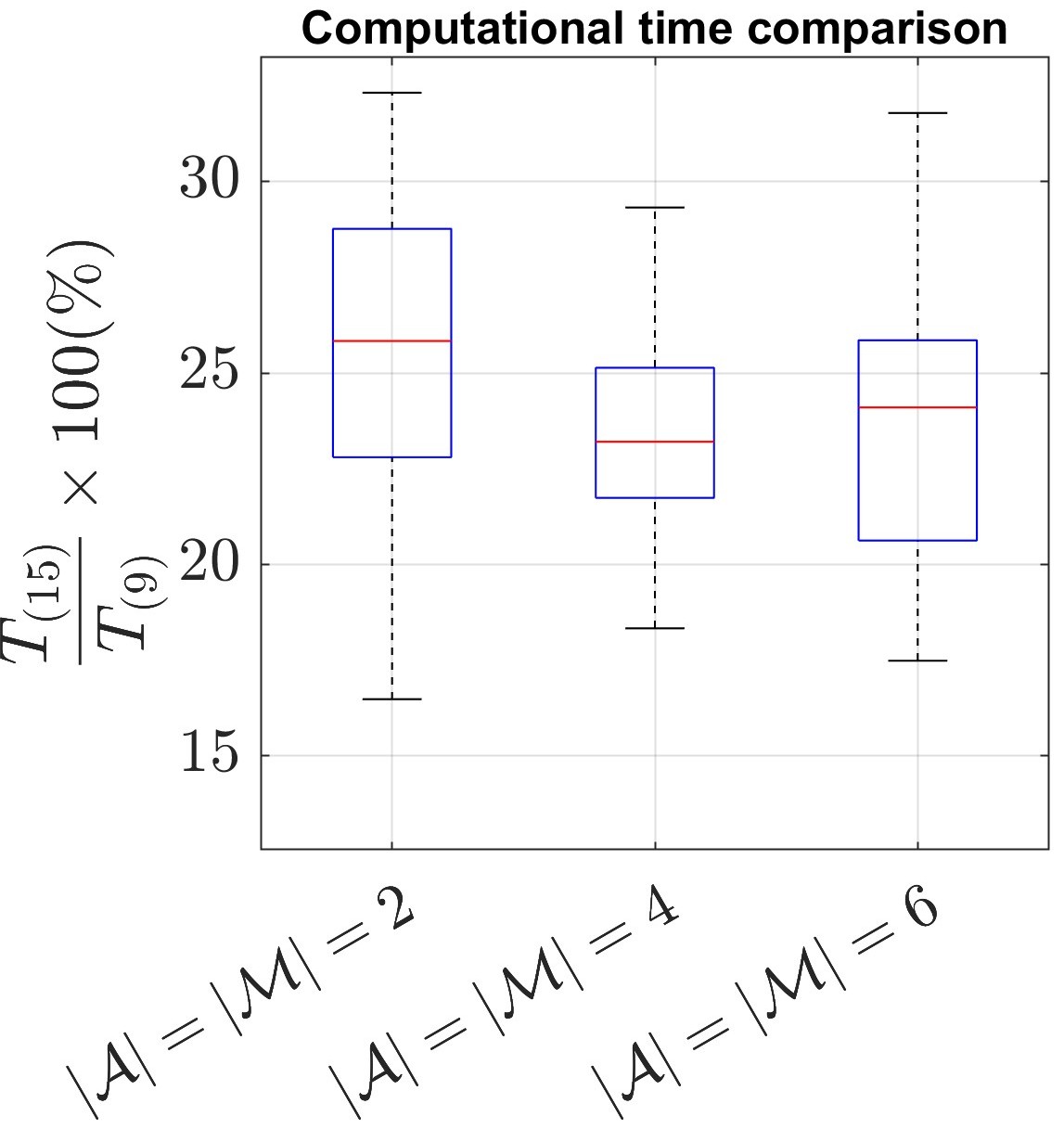}
        \caption{}
        \label{fig:SEgrid_time}
    \end{subfigure}
    \caption{a) relative comparison between the solutions to \eqref{Q_dual_min} and \eqref{Q_hat_min}, denoted as $Q$ and $\hat Q$, respectively; b) comparison in computational time of solving \eqref{Q_dual_min} and \eqref{Q_hat_min}, denoted as $T_{\eqref{Q_dual_min}}$ and $T_{\eqref{Q_hat_min}}$, respectively.}
    \label{fig:compare}
\end{figure}

The comparison between solving \eqref{Q_dual_min} and solving its proxy \eqref{Q_hat_min} is presented in Fig.~\ref{fig:compare}, which supports the theoretical finding in Theorem~\ref{th:upper_bound}.  
From Fig.~\ref{fig:SEgrid_proxy}, we observe that the relative error between the worst-case impact of stealthy attacks \eqref{Q_dual_min} and its proxy \eqref{Q_hat_min} remains under 15\% in the median for the three attack scenarios. 
Regarding the computational time, let us denote $T_{\eqref{Q_dual_min}}(n)$ and $T_{\eqref{Q_hat_min}}(n)$ as the computational times for solving \eqref{Q_dual_min} and \eqref{Q_hat_min}, respectively, where $n$ is the number of attack/monitor buses. The simulation results yield the following ranges: $T_{\eqref{Q_dual_min}}(2) \in [2.44, 4.73]$(h), $T_{\eqref{Q_dual_min}}(4) \in [2.72, 4.70]$(h), $T_{\eqref{Q_dual_min}}(6) \in [2.82, 4.62]$(h), $T_{\eqref{Q_hat_min}}(2) \in [0.65, 1.15]$(h), $T_{\eqref{Q_hat_min}}(4) \in $ \\ 
$[0.73, 1.08]$(h), and
$T_{\eqref{Q_hat_min}}(6) \in [0.75, 1.02]$(h). These results show that solving \eqref{Q_hat_min} consistently reduces the computational time by approximately 75\% in the median compared to solving \eqref{Q_dual_min}, across all attack scenarios, further demonstrating the significant computational advantage of the proposed framework.

\vspace{-0.2cm}
\section{Conclusion}\label{sec:con}
\vspace{-0.2cm}
In this paper, we proposed a metric to quantify the security of NCS when a part of the system is uncertain. We provided a computationally efficient method to determine the value of the metric when the maximum $\Lc_2$ gain of the uncertain part is available. Moreover, we discussed model-based and data-based methods to obtain such a maximum $\Lc_2$ gain. 
The results were numerically illustrated on a power transmission grid spanning Sweden and Northern Denmark. 
Future work includes addressing the scalability aspects and extension to nonlinear systems.

\bibliography{mybibfile}             

\vspace{-0.5cm}
\appendix
\section{}    
\label{app:system_parameters}
\vspace{-0.5cm}
\begin{align}
    &\bar A = \ba{cc}
    A_c & 0 \\ 0 & A_u \ea + \ba{cc}
    0 & B_c \\ B_u & 0 \ea \bar D  \ba{cc}
    C_c & 0 \\ 0 & C_u \ea, \non   \\
    &\bar F = \ba{c}
    F_x \\ 0 \ea + \ba{cc}
    0 & B_c \\ B_u & 0 \ea \bar D \ba{c} 
    F_c \\ 0 \ea, \non  \\
    &\bar C_{r} = \ba{cc} C_r & 0 \ea + \ba{cc}
    0 & D_r \ea \bar D \ba{cc}
    C_c & 0 \\ 0 & C_u \ea,  \non  \\
    &\bar F_{r} = F_r + \ba{cc}
    0 & D_r \ea \bar D \ba{c} 
    F_c \\ 0 \ea, \non  \\
    &\bar C_{p} = \ba{cc} C_p & 0 \ea + \ba{cc}
    0 & D_p \ea \bar D \ba{cc}
    C_c & 0 \\ 0 & C_u \ea, \non  \\
    &\bar F_{p} = F_p + \ba{cc}
    0 & D_p \ea \bar D \ba{c} 
    F_c \\ 0 \ea, \non \\
    & \bar D = \Bigg( I - \ba{cc}
    0 & D_c \\ D_u & 0 \ea \Bigg)^{-1}. \label{matrix_agg}
\end{align}
\begin{align}
    &R_c(\Sigma_c, P_c, \gamma_u, \gamma, \psi, \theta)  \triangleq 
    \ba{ccc}
    A_c^\top P_c + P_c A_c  ~&~ P_c B_c ~&~ P_c F_x \\ 
    B_c^\top P_c & 0 & 0 \\
    F_x^\top P_c & 0 & 0  
    \ea   \non \\
    &
    -\gamma \ba{c}
    C_r^\top \\ D_r^\top \\ F_r^\top \ea 
    \ba{c}
    C_r^\top \\ D_r^\top \\ F_r^\top 
    \ea^\top 
    - \psi \ba{c}
    0 \\ 0 \\ I 
    \ea \ba{c}
    0 \\ 0 \\ I 
    \ea^\top - \theta \ba{c} 
    0 \\ I \\ 0 
    \ea \ba{c}
    0 \\ I \\ 0 
    \ea^\top \non \\
    & + \ba{c}
    C_p^\top \\ D_p^\top \\ F_p^\top 
    \ea \ba{c}
    C_p^\top \\ D_p^\top \\ F_p^\top 
    \ea^\top 
    + \theta \gamma_u 
    \ba{c}
    C_c^\top \\ D_c^\top \\ F_c^\top 
    \ea \ba{c}
    C_c^\top \\ D_c^\top \\ F_c^\top 
    \ea^\top. \label{matrix_R1}
\end{align}
                                                                         
\end{document}